\documentstyle[epsbox]{ioplppt}
\begin{document}
\letter{ Derivation of Boltzmann Equation\\
         in Closed-Time-Path Formalism
       }
\author { Jun Koide }
\address{ Department of Physics, 
          Faculty of Science and Technology, 
          Keio University, Yokohama 223\\
        }
\begin{abstract}
A systematic derivation of Boltzmann equation is presented
in the framework of closed-time-path formalism.
Introducing a new type of probe,
the expectation value of number operator is calculated 
as a functional of source.
Then solving for the source by inverting the relation,
the equation of motion for number is obtained
when the source is removed,
and it turns out to be the Boltzmann equation.
The inversion formula is used in the course of derivation.
\end{abstract}
\newcommand{\appref}[1]{Appendix \ref{#1}}
\newcommand{\citen}[1]{\cite{#1}}
\newcommand{\ve}[1]{ {\bf #1} }
\newcommand{\dspfrac}[2]
           {\frac{\displaystyle \strut #1}{\displaystyle{#2}}}
\newcommand{\txtfrac}[2]
           {{\textstyle \frac{#1}{#2} }}
\newcommand{\lra}[1]{\left( #1 \right)}
\newcommand{\la}[1]{\left( #1 \right.}
\newcommand{\ra}[1]{\left. #1 \right)}
\newcommand{\lrb}[1]{\left\{\, #1 \,\right\}}
\newcommand{\lb}[1]{\left\{\, #1 \,\right.}
\newcommand{\rb}[1]{\left.\, #1 \,\right\}}
\newcommand{\lrc}[1]{\left[\, #1 \,\right]}
\newcommand{\lc}[1]{\left[\, #1 \,\right.}
\newcommand{\rc}[1]{\left.\, #1 \,\right]}
\newcommand{\lrangle}[1]{\left<\, #1 \,\right>}
\newcommand{\lrang}[1]{\left<\, #1 \,\right>}
\newcommand{\lang}[1]{\left<\, #1 \,\right.}
\newcommand{\rang}[1]{\left.\, #1 \,\right>}
\newcommand{\lrabs}[1]{\left|\, #1 \,\right|}
\newcommand{\rbar}[2]{\left.\, {#1} \,\right|_{#2}}

\newcommand{\re}{{\rm e}}
\newcommand{\ri}{{\rm i}}
\newcommand{\rO}{{\rm O}}
\newcommand{\fdag}{ \mbox{\footnotesize \dag} }

\newcommand{\vk}{\ve{k}}
\newcommand{\vq}{\ve{q}}
\newcommand{\vkk}{\ve{k}'}
\newcommand{\vqq}{\ve{q}'}

\newcommand{\rd}{{\rm d}}
\newcommand{\del}{\partial}
\newcommand{\Del}{{\mit \Delta}}
\newcommand{\rDel}{{\rm \Delta}}
\newcommand{\delt}{\partial_{t}}

\newcommand{\cD}{{\cal D}} 
\newcommand{\cM}{{\cal M}} 

\newcommand{\rI}{{\rm I}}
\newcommand{\rF}{{\rm F}}
\newcommand{\rC}{{\rm C}}
\newcommand{\lx}{{\rm X}}

\newcommand{\tI}{t_{\rI}}
\newcommand{\tF}{t_{\rF}}
\newcommand{\tint}{{\int_{\tI}^{\tF}\hspace{-0.3em}}} 
\newcommand{\tintI}{{\int_{\tI}^{t}\hspace{-0.3em}}} 
\newcommand{\intst}{{\int_{s}^{t}\hspace{-0.2em}}} 
\newcommand{\bint}{{\int_{0}^{\beta \hbar}\hspace{-0.3em}}} 

\newcommand{\cnt}{ \rC }

\newcommand{\psiD}{\psi_{\rDel}}
\newcommand{\psiC}{\psi_{\rC}}
\newcommand{\opsi}{\hat{\psi}}
\newcommand{\psid}{\hat{\psi}^{\fdag}}

\newcommand{\nk}{n_{\vk}}
\newcommand{\nq}{n_{\vq}}
\newcommand{\nqq}{n_{\vqq}}
\newcommand{\nqqk}{n_{\vq+\vqq-\vk}}
\newcommand{\no}{n^{(0)}}
\newcommand{\nok}{n^{(0)}_{\vk}}
\newcommand{\noq}{n^{(0)}_{\vq}}
\newcommand{\noqq}{n^{(0)}_{\vqq}}
\newcommand{\noqqk}{n^{(0)}_{\vq+\vqq-\vk}}
\newcommand{\epk}{\epsilon_{\vk}}
\newcommand{\epq}{\epsilon_{\vq}}
\newcommand{\epqq}{\epsilon_{\vqq}}
\newcommand{\epqqk}{\epsilon_{\vq+\vqq-\vk}}
\newcommand{\Jk}{J{\vk}}
\newcommand{\Jq}{J{\vq}}
\newcommand{\Jqq}{J{\vqq}}
\newcommand{\Jqqk}{J{\vq+\vqq-\vk}}

\newcommand{\hast}{h^{\ast}}
\newcommand{\kast}{k^{\ast}}
\newcommand{\ho}{h_{0}}
\newcommand{\ko}{k_{0}}
\newcommand{\hoast}{h_{0}^{\ast}}
\newcommand{\koast}{k_{0}^{\ast}}

\newcommand{\GA}{G^{\rm A}}
\newcommand{\GR}{G^{\rm R}}
\newcommand{\GAFI}{G^{\rm A}_{\rF \rI}}
\newcommand{\GRIIF}{G^{\rm R}_{\rI ' \rF}}

\newcommand{\Jd}{J_{\rDel}} 
\newcommand{\Jc}{J_{\cnt}} 
\newcommand{\Jdd}{J_{\rDel}^{\ast}} 
\newcommand{\Jcd}{J_{\cnt}^{\ast}} 
\newcommand{\Jdi}{\mathop{J_{\rDel ,i}}\nolimits}
\newcommand{\Jci}{\mathop{J_{\rC ,i}}\nolimits}
\newcommand{\Jdj}{\mathop{J_{\rDel ,j}}\nolimits}
\newcommand{\Jcj}{\mathop{J_{\rC ,j}}\nolimits}

\newcommand{\mA}{m_{A}} 
\newcommand{\mB}{m_{B}} 
\newcommand{\epsAk}{\epsilon^{A}_{\vk}} 
\newcommand{\epsBk}{\epsilon^{B}_{\vk}} 
\newcommand{\egap}{\varepsilon_{{\rm bind}}} 

\newcommand{\muD}{\mu_{\rDel}} 
\newcommand{\muC}{\mu_{\cnt}} 
\newcommand{\muAD}{\mu^{A}_{\rDel}} 
\newcommand{\muAC}{\mu^{A}_{\cnt}} 
\newcommand{\muBD}{\mu^{B}_{\rDel}} 
\newcommand{\muBC}{\mu^{B}_{\cnt}} 

\noindent
In this letter,
we present a new approach to derive the Boltzmann equation (BE).
There have been some works on this subject
both in the framework of closed-time-path (CTP) formalism,%
~\cite{Lawrie,Niegawa}
and in the framework of thermo-field dynamics.~\cite{Umezawa}
%
These approaches have an advantage
that the time dependence of the number is not introduced by hand.
Instead,
a counter-term is first introduced 
into the CTP or thermo-field Lagrangian 
and bare propagator is calculated.
Then to determine the counter-term,
some condition,
such as the cancelation of on-shell part 
of the self-energy~\cite{Lawrie,Umezawa}
or the cancelation of pinch-singularity,~\cite{Niegawa} 
is adopted,
which leads to BE.
Since their primal purpose 
is to construct the non-equilibrium perturbation theory,
BE appears as a byproduct.
It is preferable if we can derive BE more directly
as an equation of motion (EoM) 
of expectation value of the number operator.
Moreover in their approaches,
the condition to determine the counter-term is of course not unique,
and the approximation made is not so clear.
In the following,
a more direct approach is studied
based on the inversion method.~\cite{Fukuda,suppl}
A new type of probe dictated from the counter-term approach 
is introduced in course of derivation.

Let us briefly describe the inversion method 
which is a systematic procedure to derive EoM in CTP formalism.
In CTP formalism,~\cite{CTP,Chou}
we introduce a time dependent source $J$ 
to probe some operator of interest,
say $Q(\hat{\varphi})$, 
which is a function of the dynamical variable $\hat{\varphi}$.
Then with the Hamiltonian $\hat{H}$ of $\hat{\varphi}$,
the CTP generating functional is defined as
\begin{eqnarray}
 \fl
 \hspace{2em}
   \re^{ \frac{\ri}{\hbar}W \lrc{J_{1},J_{2}} } 
    &\equiv \Tr \, 
               T \,
               \re^{-\frac{i}{\hbar} \tint \rd t \,
                     \lra{ \hat{H}-J_{1}(t)\hat{Q} }
                   }
               \hat{\rho}\,\,
               \tilde{ T }\,
               \re^{ \frac{i}{\hbar} \tint \rd t \, 
                     \lra{ \hat{H}-J_{2}(t)\hat{Q} }
                   } 
\label{W1} \\
\fl
    &\propto 
        \int \lrc{ \rd \varphi_{1} \,d{\varphi_{2}}  }
        \lrangle{ {\varphi_{1}}_{\rI} 
                  \lrabs{\hat{\rho}}
                  {\varphi_{2}}_{\rI}
                }
         \re^{ \frac{i}{\hbar} \tint \rd t\, 
               \lra{ L(\varphi_{1}) - L(\varphi_{2})
                    +J_{1}Q(\varphi_{1})-J_{2}Q(\varphi_{2}) }
             }
,
\label{W2} 
\end{eqnarray}
where $\hat{\rho}$ is the initial distribution
and $T$ and $\tilde{T}$ are 
the time ordering and anti-ordering operators,
respectively.
The last equality is due to path-integral representation,
where $\varphi_{1}$ and $\varphi_{2}$ are respectively introduced 
as integral variables along the forward and backward time branches.

For convenience of later discussion,
let us introduce ``physical'' representation~\cite{Chou}
through
$
  \Jc \equiv \frac{1}{2}(J_{1}+J_{2})
$,
and
$
  \Jd \equiv J_{1}-J_{2} 
$.
Then $ \Jd = 0 $ is physical
and $\Jc$ plays the role of external force. 
The expectation value of $\hat{Q}$ at time $t$
under physical external source $\Jc=J$ can be calculated as
\begin{equation}
  Q(t) 
    \equiv \rbar{ \frac{\delta W \lrc{ \Jd,\Jc }}{\delta \Jd (t)} }
                {\stackrel{\scriptstyle \Jc=J}{\Jd=0}}
    = \lrangle{ \hat{Q}(t) }_{J}
.
\label{Wd0}
\end{equation}
This gives us the expectation value $Q$ 
as a functional of external source $J$.

In order to obtain the EoM of $Q$,
we solve the relation (\ref{Wd0}) inversely 
to express $J$ as a functional of $Q$.
Then setting the external source $J=0$,
the obtained relation gives EoM of $Q$.
(Inversion method~\cite{suppl})
Formally,
the general expression of EoM can be written 
with the Legendre transformation of $W$.
But practically,
the process of Legendre transformation is unnecessary,
and in this letter, 
this inversion is carried out 
in the following perturbative fashion.

Usually $Q$ as a functional of $J$ 
is obtained as some perturbation series 
\begin{equation}
  Q(t) = f\lrc{t;J} = \sum_{n}\varepsilon^{n}f^{(n)}\!\lrc{t;J}
,
\label{f-ser}
\end{equation}
where $\varepsilon$ is a small parameter
and $f\lrc{t;J}$ expresses 
that $f$ is a function of $t$ and functional of $J$.
Then if we write the inverted relation as
\begin{equation}
  J(t) = g\lrc{t;Q} = \sum_{m}\varepsilon^{m}g^{(m)}\!\lrc{t;Q}
,
\label{g-ser}
\end{equation}
following simple identity is obtained
\begin{eqnarray}
\fl
  Q(t) 
    &= f\lrc{t;g\lrc{Q}}
\nonumber\\
\fl &= f^{(0)}\!\lrc{t;g^{(0)}\!\lrc{Q}}
     +\varepsilon 
      \lra{ \int\!\rd s\frac{\delta f^{(0)}(t)}{\delta g^{(0)}(s)}
               g^{(1)}\!\lrc{s;Q}
           +f^{(1)}\!\lrc{t;g^{(0)}\!\lrc{Q}}
           }
\nonumber\\
\fl &\hspace{1em}
     +\varepsilon^{2}
      \la{  \int\!\rd s \frac{\delta f^{(0)}(t)}{\delta g^{(0)}(s)}
               g^{(2)}\!\lrc{s;Q}
           +\frac{1}{2}\int\! \rd s\, \rd s'
               \frac{\delta^{2} f^{(0)}(t)}
                    {\delta g^{(0)}(s)\delta g^{(0)}(s')}
               g^{(1)}\!\lrc{s;Q} g^{(1)}\!\lrc{s';Q}
          }
\nonumber\\
\fl &\hspace{4em}
      \ra{ +\int\!\rd s\frac{\delta f^{(1)}(t)}{\delta g^{(0)}(s)}
               g^{(1)}\!\lrc{s;Q}
           +f^{(2)}\!\lrc{t;g^{(0)}\!\lrc{Q}}
         }
      +\rO(\varepsilon^{3})
,
\end{eqnarray}
where,
e.g. $\delta f^{(0)}[t;J]/\delta J(s)$ evaluated at $J=g^{(0)}[Q]$ 
is abbreviated as $\delta f^{(0)}(t)/\delta g^{(0)}(s)$.
Comparing the lhs and rhs in each order of $\varepsilon$,
we obtain the expressions for $g^{(m)}$ in terms of $f^{(n)}$,
which we call the ``inversion formulae''.~\cite{suppl}
\begin{eqnarray}
\fl
  g^{(0)}\!\lrc{t;Q}
     =& {f^{(0)}}^{-1}\!\lrc{t;Q}
,
\label{inv0}
\\
\fl
  g^{(1)}\!\lrc{t;Q}
     =& -\!\int\!\rd t'
         \lra{ \frac{\delta f^{(0)}}{\delta g^{(0)}} 
             }^{-1}\hspace{-1.2em}(t,t')\:
         f^{(1)}\![t';g^{(0)}]
,
\label{inv1}
\\
\fl
  g^{(2)}\!\lrc{t;Q}
     =& -\!\int\!\rd t'
         \lra{ \frac{\delta f^{(0)}}{\delta g^{(0)}} 
             }^{-1}\hspace{-1.2em}(t,t')\:
         \la{  \frac{1}{2}\int\! \rd s\, \rd s'
                  \frac{\delta^{2} f^{(0)}(t')}
                       {\delta g^{(0)}(s)\delta g^{(0)}(s')}
                  g^{(1)}\!\lrc{s;Q} g^{(1)}\!\lrc{s';Q}
            }
\nonumber\\
\fl   &\hspace{11em}
         \ra{ +\int\!\rd s\frac{\delta f^{(1)}(t')}{\delta g^{(0)}(s)}
                  g^{(1)}\!\lrc{s;Q}
              +f^{(2)}\![t';g^{(0)}]
            }
.
\label{inv2}
\end{eqnarray}

First of all, 
to make this method work,
we need a non-trivial lowest-order functional expression $f^{(0)}[t;J]$ 
which can be inversely solved for $J$.
This becomes the key-point for deriving BE.
If we naively apply this method to number operator,
the expectation value does not reveal 
such non-trivial dependence on $J$.

Let us see the problem more closely.
We consider a non-relativistic Boson field
of a homogeneous system
described by the Hamiltonian
$
 H = H_{0}+H_{\rm int}
$
with 
$
 H_{0} = \sum_{\vk} \epsilon_{\vk} \opsi^{\fdag}_{\vk} \opsi_{\vk} 
$,
and 
$
 H_{\rm int} = \frac{\lambda}{4}\sum_{\vk,\vkk,\vq} 
               \opsi^{\fdag}_{\vk+\vq} \opsi^{\fdag}_{\vkk-\vq} 
               \opsi_{\vk} \opsi_{\vkk} 
$,
where $\lambda$ is a coupling constant,
which is assumed to be small 
and plays the role of $\varepsilon$ in (\ref{f-ser}).
Extension to other type of interaction is straightforward.
For the initial density matrix $\hat{\rho}$,
we assume that 
no initial correlation exists 
among the different wave-number components 
and $\hat{\rho}$ can be written
as a product form $\prod_{\vk}\hat{\rho}_{\vk}$,
where $\hat{\rho}_{\vk}$ is a density for each wave-number 
which gives 
$
 n_{\vk}(\tI) 
   = \Tr \hat{\rho}_{\vk}\opsi^{\fdag}_{\vk} \opsi_{\vk}
$.

In order to derive EoM of the expectation value of the number 
$
 \hat{n}_{\vk}(t)=\opsi^{\fdag}_{\vk}(t)\opsi_{\vk}(t)
$,
a naive choice of the source is 
$
  \hat{H}-\sum_{\vk}J_{\vk}(t) \opsi^{\fdag}_{\vk}(t)\opsi_{\vk}(t)
$.
Then in path-integral representation of CTP generating functional,
this source can be built into the free part of the Lagrangian as 
\begin{equation}
  L^{J}_{0}(\psi_{1})-L^{J}_{0}(\psi_{2})
    = \sum_{\vk} 
        \psi^{\ast}_{i,\vk} \cD_{ij,\vk} \psi_{j,\vk}
\end{equation}
with the matrix
\begin{equation}
  \cD_{\vk}(t,\delt)
    \equiv
      \lra{ \begin{array}{cc}
               \ri \hbar \delt -\epsilon_{\vk} + J_{\vk}(t)     & 0  
            \\
               0     & -\ri \hbar \delt +\epsilon_{\vk} - J_{\vk}(t) 
            \\
            \end{array}
          }
.
\nonumber\\
\label{L0}
\end{equation}
The bare propagator is essentially 
the inverse of the matrix in (\ref{L0}),
and with this propagator,
if we evaluate the expectation value $n_{\vk}(t)$ 
in the absence of interaction,
we just obtain the initial value
$
  \langle \hat{n}_{\vk}(t) \rangle_{J} = n_{\vk}(\tI)
$,
due to the conservation of $\hat{n}_{\vk}$ 
for $\lambda =0$ even when $J_{\vk}\neq 0$.
Since no dependence on $J$ appears,
we fail to obtain the inversion in lowest-order,
and hence the inversion formulae can not be used in this case.
Probe of the form (\ref{L0}) is not enough 
to handle the number operator.

Then why does the counter-term method work?
According to reference~\citen{Lawrie},
the time-local counter-term is constructed 
so as to keep the following structure 
of the full propagator in CTP formalism 
(We suppress the index of wavenumber for a while.)
\begin{eqnarray}
 \lo{G(t,s)}
  &\equiv
        -\Tr \, \hat{\rho}
         \lra{ \begin{array}{cc}
                T \opsi(t)\psid(s) &          \psid(s)\opsi(t) 
               \\
                \opsi(t)\psid(s)   & \tilde{T}\opsi(t)\psid(s)
               \end{array}
             }_{\rm c}
\nonumber \\
  &=  \theta(t-s) \lra{\begin{array}{cc}
                         h(t,s) & k(t,s) \\
                         h(t,s) & k(t,s) \\
                       \end{array}
                      }
     +\theta(s-t) \lra{\begin{array}{cc}
                         k^{\ast}(s,t) & k^{\ast}(s,t) \\
                         h^{\ast}(s,t) & h^{\ast}(s,t) \\
                       \end{array}
                      }
,
\label{structure1}
\end{eqnarray}
where `c' means the connected part and
\begin{equation}
  h(t,s) \equiv -\langle \opsi(t)\psid(s) \rangle_{\rm c}
,
\label{hdef}
\hspace{2em}
  k(t,s) \equiv -\langle \psid(s)\opsi(t) \rangle_{\rm c}
.
\label{kdef}
\end{equation}
Then it turns out
that the counter-term $\psi^{\ast}_{i}\cM_{ij}\psi_{j}$ 
with the matrix 
\begin{equation}
  \cM(t) 
   = \lra{\begin{array}{cc}
              \hbar \Del \omega(t) -\ri \alpha(t) 
           & -\ri(\hbar \gamma(t) -\alpha(t))
          \\
              \ri(\hbar \gamma(t)+\alpha(t))
           & -\hbar \Del \omega(t) -\ri \alpha(t)
          \end{array}
         }
,
\label{counter}
\end{equation}
is allowed to be subtracted from the free part of the Lagrangian,
where $\Del \omega$, $\alpha$ and $\gamma$ are all real functions
which are determined by appropriate conditions.
The bare propagator calculated from 
$L_{0}(\psi_{1})-L_{0}(\psi_{2})- \psi^{\ast}_{i}\cM_{ij}\psi_{j}$
leads to non-trivial time dependence of the number 
in the absence of interaction.
The existence of the parameters in non-diagonal elements
is a crucial point.

Comparing (\ref{counter}) with (\ref{L0}),
the parameter we have utilized 
as a physical external source in (\ref{L0}) 
corresponds to $\Del \omega$ in (\ref{counter}).
Equation (\ref{counter}),
however,
suggests that another physical source
corresponding to $\alpha$ or $\gamma$ can be introduced as a probe.
Our choice here is the source corresponding to $\alpha$.
(The source corresponding to $\gamma$ can be treated similarly.)
Then the free part of the Lagrangian including the source 
now has the matrix
\begin{equation}
  \cD(t,\delt)  
   = \lra{\begin{array}{cc}
              \ri \hbar \delt -\epsilon +\ri J(t)       & -\ri J(t)
           \\
             -\ri J(t)       & -\ri \hbar \delt +\epsilon +\ri J(t)
           \\
           \end{array}
          }
.
\label{D1}
\end{equation}
Note that although the source $J$ is introduced in this way,
what we calculate in the following is just the expectation value of the number;
We integrate $\psi^{\ast}_{1}(t+\varepsilon)\psi_{1}(t)$
under the existence of the probe (\ref{D1}).
(Of course other choices,
 e.g. $\psi^{\ast}_{2}(t-\varepsilon)\psi_{2}(t)$,
 produce the same results.)

From the matrix (\ref{D1}),
the bare propagator $G_{0}$ is calculated by
\begin{eqnarray}
 \cD (t,\delt) G_{0}(t,s) 
   &=& G_{0}(t,s) \cD (s,-\!\!\stackrel{\leftharpoonup}{\del}_{s}) 
\\
   &=& -\ri \hbar \delta(t-s) 
.
\label{DGinv}
\end{eqnarray}
Since $\cD$ has been chosen 
as to keep the structure (\ref{structure1}) unchanged,
$G_{0}$ has the same structure
in which $h$ and $k$ are replaced by $h_{0}$ and $k_{0}$,
respectively.
Then (\ref{DGinv}) leads to the equations
\begin{eqnarray}
  \lra{\ri \hbar \delt - \epsilon }h_{0}(t,s) &=& 0
,
\label{hoko1}\\
  \lra{\ri \hbar \delt - \epsilon }k_{0}(t,s) &=& 0
,
\label{hoko2}
\end{eqnarray}
for $t>s$, 
and
\begin{eqnarray}
  \lra{\ri \hbar \delt -\epsilon +\ri J(t)}\koast(s,t) 
    &=& \ri J(t) \hoast(s,t)
,
\label{hoko3}\\
  \lra{\ri \hbar \delt -\epsilon -\ri J(t)}\hoast(s,t) 
    &=&-\ri J(t) \koast(s,t)
,
\label{hoko4}
\end{eqnarray}
for $s>t$.
The boundary conditions at $t=s$ are given as
\begin{eqnarray}
  h_{0}(s,s) - \koast(s,s) = -1
,
&\hspace{2em}&
  k_{0}(s,s) - \hoast(s,s) = 1
,
\label{cond12}\\
  h_{0}(s,s) - \hoast(s,s) = 0
,
&\hspace{2em}&
  k_{0}(s,s) - \koast(s,s) = 0
.
\label{cond34}
\end{eqnarray}
From (\ref{cond34})
$h_{0}(s,s)$ and $k_{0}(s,s)$ are real functions.
Then the two conditions in (\ref{cond12}) are identical 
and simply express the fact that 
the expectation value 
of the equal-time commutator $[\opsi,\psid]$ is unity.
Note that from the definition (\ref{kdef}), 
$k_{0}(t,t)$ gives the expectation value 
of the number operator 
(multiplied by $-1$) in the absence of the interaction.
which we denote as $\no(t)$.

From (\ref{hoko1}) and (\ref{hoko2}),
we obtain for $t>s$ 
\begin{eqnarray}
  k_{0}(t,s) 
    &=  \re^{-\frac{\ri}{\hbar} \epsilon (t-s)} k_{0}(s,s)
    &=   -\no(s) \: \re^{-\frac{\ri}{\hbar} \epsilon (t-s)} 
,
\label{k1}
\\
  h_{0}(t,s) 
    &=  \re^{-\frac{\ri}{\hbar} \epsilon (t-s)} h_{0}(s,s)
    &=   -(\no(s)+1)\: \re^{-\frac{\ri}{\hbar} \epsilon (t-s)}
.
\label{h1}
\end{eqnarray}
Then exchanging $t$ and $s$ in (\ref{k1}) and (\ref{h1})
and taking the complex conjugation,  
$\hoast(s,t)$ and $\koast(s,t)$ are obtained for $s>t$.
Substituting them into (\ref{hoko3}) or (\ref{hoko4}),
both equations turn out to give an identical result,
and we find that $\no$ must satisfy the condition 
\begin{equation}
  J(t) = \hbar \delt  \no(t) 
,
\label{no-eom}
\end{equation}
which gives EoM for $\no$ and is integrated as
\begin{equation}
  \no\lrc{t;J} = \no(\tI)+\tintI \rd s \frac{J(s)}{\hbar}
.
\label{n-lowest}
\end{equation}
Equations (\ref{k1}),(\ref{h1}) and (\ref{n-lowest}) 
together with the structure like (\ref{structure1})
determine the bare propagator.

As already seen from (\ref{n-lowest}) or (\ref{no-eom}),
we succeeded to make the expectation value of the number 
depend on $J$ in the lowest order,
i.e. O($\lambda^{0}$).
This makes the inversion formula applicable.
The rhs of equation (\ref{n-lowest}) corresponds 
to the desired lowest-order functional $f^{(0)}$ in (\ref{f-ser}),
and (\ref{no-eom}) is the inverted relation,
the rhs of which corresponds to $g^{(0)}$ of (\ref{g-ser}).
So our next task is to calculate the perturbative correction to $n$,
and then to derive the correction to the EoM (\ref{no-eom})
with the aid of the inversion formulae.

\begin{figure}
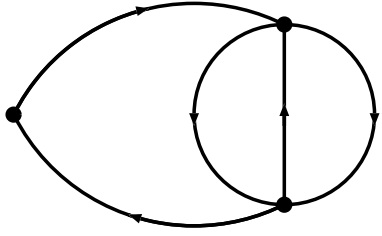

\hspace{\mathindent}
\epsfile{file=fig1.eps,width=5cm}
\caption{Diagram for $O(\lambda^{2})$-correction to $n_{\vk}$.}
\label{n2}
\end{figure}
With the propagator calculated above,
the non-zero perturbative correction to $n_{\vk}(t)$
first comes from a diagram shown in \fref{n2},
which is of $\rO(\lambda^{2})$.
The contributions of $\rO(\lambda)$
from a tadpole type self-energy insertion vanishes 
due to the cancelation of terms 
from the vertices on forward and backward time branches, 
and similarly for the two tadpoles inserted diagram 
of $\rO(\lambda^{2})$.
As the result
\begin{eqnarray}
\fl
  \nk\lrc{t,J} = \nok(t)
     +\lra{\frac{\lambda}{\hbar}}^{2} \sum_{\vq,\vqq}
     \tintI \rd t' \int_{\tI}^{t'} \hspace{-0.3em}\rd s'
      \cos \lra{\omega_{\vk,\vq,\vqq}(t'-s')}
\nonumber \\
\fl \hspace{1.5em}
     \times
      \lrb{ (\nok+1) (\noqqk+1) \noq \noqq
           -\nok \noqqk (\noq+1) (\noqq+1) 
          }(s')
.
\label{nAC2}
\end{eqnarray}
where 
$
  \omega_{\vk,\vq,\vqq}
    \equiv \frac{1}{\hbar}(\epq+\epqq-\epqqk-\epk)
$.
Recall that all $\nok$'s are functionals of $J_{\vk}$ 
given in (\ref{n-lowest}).
Equation (\ref{nAC2}) corresponds 
to $f^{(0)}+\varepsilon f^{(1)}+\varepsilon^{2} f^{(2)}$ 
of (\ref{f-ser}),
where $f^{(1)}$ vanishes as mentioned above.

Applying the inversion formulae,
we obtain the correction to (\ref{no-eom}) as
\begin{eqnarray}
\fl
  J_{\vk}(t) 
   =& \hbar \delt \nk(t)
      -\frac{\lambda^{2}}{\hbar}\sum_{\vq,\vqq}
       \tintI \rd s
       \cos \lra{\omega_{\vk,\vq,\vqq}(t-s)}
\nonumber\\
\fl & \times
      \lrb{ (\nk+1) (\nqqk+1) \nq \nqq 
           -\nk \nqqk (\nq+1) (\nqq+1) }(s)
.
\label{muC2}
\end{eqnarray}
Note that,
in course of the inversion,
all the functionals of $J$ 
are evaluated at $J_{\vk}=\hbar \dot{n}_{\vk}$
and $\nok\lrc{t;J}$ contained therein becomes $n_{\vk}(t)$. 
If we set the external source $J=0$,
EoM for the number is obtained.
The correction term is similar 
to the collision terms of Boltzmann equation,
but it has non-Markovian form 
and contain energy non-conserving process.

The ordinary Markovian BE
is obtained by the adiabatic expansion.
Setting the initial time $\tI =-\infty$,
we abbreviate the products of $n$ and $n+1$ 
in the integrand of (\ref{muC2}) as $N^{(2)}$ 
and expand it around the time $t$ as 
$
  N^{(2)}(s)=N^{(2)}(t)+(s-t)\dot{N}^{(2)}(t)+\cdots
$,
regarding the time differentiations to be small.
Then the integral becomes
\begin{equation}
\fl
  \int_{-\infty}^{t} \rd s 
    \cos \omega(t-s) N^{(2)}(s)
  =  \pi \delta(\omega)N^{(2)}(t)
     +\frac{\wp}{\omega^{2}} \dot{N}^{(2)}(t)
     +\cdots
.
\label{adiabatic}
\end{equation}
The second term is proportional to $\dot{n}$ 
and gives a perturbative correction 
to the coefficient of the first term in rhs of (\ref{muC2}),
which can be neglected.
Regarding all higher time derivatives to be small,
we take into account up to the first term of (\ref{adiabatic}),
and obtain the ordinary time-local BE with energy conserving process 
\begin{eqnarray}
\fl
  \hbar \delt \nk(t)
   =& \pi\lambda^{2}\sum_{\vq,\vqq}
       \delta \lra{\epq+\epqq-\epqqk-\epk}
\nonumber\\
\fl & \times
      \lrb{ (\nk+1) (\nqqk+1) \nq \nqq 
           -\nk \nqqk (\nq+1) (\nqq+1) }(t)
,
\label{BE}
\end{eqnarray}

The key-point of our derivation 
is the new type of probe introduced in (\ref{D1}).
After that,
the application of inversion formula is straightforward.
Of course 
with the usage of higher-order inversion formulae,~\cite{suppl}
we can calculate  higher-order corrections to EoM quite systematically.
This will be presented in other place.

Physical contents of (\ref{D1}) becomes somewhat clear
if we consider effective action of $\psi$.
From the CTP generating functional $W$ 
with $\psi$ itself as the order parameter $Q$,
the effective action $\Gamma\lrc{\psiD,\psiC}$ is calculated
through the Legendre transformation of $W$,
where $\psiD \equiv \delta W/\delta \Jc$ and
$\psiC \equiv \delta W/\delta \Jd$.
Roughly speaking,
$\psiD=\psi_{1}-\psi_{2}$, 
$\psiC=\frac{1}{2}(\psi_{1}+\psi_{2})$
and $\cD$ is the tree part of second derivative of $\Gamma$.
Then the source of the form (\ref{D1}) 
couples to $\psi^{\ast}_{\rDel}\psiD$
and corresponds to the quantity 
$
  {\delta^{2}\Gamma}/{\delta \psi^{\ast}_{\rDel} \delta \psiD}
$
which is the 1-particle-irreducible amputated part 
of the correlation function $ \langle \{ \psid,\opsi \} \rangle $.
This may be the reason 
why we can handle the number with this source.
Another choice of the source 
corresponding to $\gamma$ in (\ref{counter}) 
also produces non-trivial time dependence in the lowest order
and EoM can be derived.
Although the result has somewhat complicated expression,
it agrees with (\ref{BE}) after the adiabatic expansion.

\vspace{1em}
\noindent
The author is grateful to professor R. Fukuda for helpful discussions.

\section*{References}
\begin{thebibliography}{99}
%
\bibitem{Lawrie}
Lawrie~I~D 1988 \JPA {\bf 21} L823 
%
\nonum
Lawrie~I~D 1989 \PR D {\bf 40} 3330 
%
\nonum
Lawrie~I~D 1992 \JPA {\bf 25} 6493
%
\nonum
Lawrie~I~D and McKernan~D~B 1997 \PR D {\bf 55} 2290
%
\bibitem{Niegawa}
Ni\'egawa~A 1998 \PL B {\bf 416} 137 
%
\nonum
Ni\'egawa~A 1999 {\it Prog.~Theor.~Phys.} {\bf 102} 1 
%
\bibitem{Umezawa}
Yamanaka~Y, Umezawa~H, Nakamura~K and Arimitsu~T 1994
{\it Int.~J.~Mod.~Phys.}~A {\bf 9} 1153 
%
\nonum
Chu~H and Umezawa~H 1994 {\it Int.~J.~Mod.~Phys.}~A {\bf 9} 1703; 2363
%
\bibitem{Fukuda}
Fukuda~R 1988 \PRL {\bf 61} 1549
%
\bibitem{suppl}
Fukuda~R, Komachiya~M, Yokojima~S, Suzuki~Y, Okumura~K and Inagaki~T 
1995 {\it Prog. Theor. Phys. Suppl.} {\bf 121} 1
%
\bibitem{CTP}
Schwinger~J 1961 \JMP {\bf 2} 407 
%
\nonum
Keldysh~L~V 1965 {\it Sov.~Phys.~-JETP} {\bf 20} 1018
%
\nonum
Landsman~N~P and van~Weert~Ch~G 1987 {\it Phys.~Rep.} {\bf 145} 141 
and references cited therein.
%
\bibitem{Chou}
Chou~K, Su~Z, Hao~B and Yu~L 1985 {\it Phys.~Rep.} {\bf 118} 1
%
\end{thebibliography}
\end{document}